\documentclass[journal,twoside,web]{ieeecolor}

\def\isPrint{1} 

\usepackage{generic}
\usepackage{cite}
\usepackage{algorithmic}
\usepackage{graphicx}
\usepackage{algorithm,algorithmic}
\usepackage{hyperref}
\hypersetup{hidelinks=true}
\usepackage{textcomp}

\usepackage{amsmath,amssymb,amsfonts,mathtools}
\usepackage{textcomp}
\usepackage{xcolor}
\usepackage{siunitx}
\usepackage{makecell}
\usepackage{array}
\usepackage{booktabs}
\usepackage{soul}
\usepackage[nolist]{acronym}
\usepackage{svg}
\usepackage{mathdots}
\usepackage{pgfplots}
\usepackage{caption}
\usepackage{subcaption}
\usepackage{booktabs}

\usepackage{tikz}
\usepackage{tikz-3dplot}
\usetikzlibrary{positioning,decorations.pathreplacing,calligraphy}

\renewcommand{\vec}{\mathbf}
\newcommand{\mat}{\mathbf}

\newcommand{\etal}{~\textit{et al.}}

\mathchardef\mhyphen="2D

\usepackage{orcidlink}

\newcommand{\orcidauthorA}{\orcidlink{0000-0002-3167-760X}} 
\newcommand{\orcidauthorB}{\orcidlink{0000-0003-1460-1061}} 
\newcommand{\orcidauthorC}{\orcidlink{0000-0002-5304-4213}} 
\newcommand{\orcidauthorH}{\orcidlink{0000-0001-5877-1439}} 
\newcommand{\orcidauthorL}{\orcidlink{0000-0002-4806-9838}} 
\newcommand{\orcidauthorM}{\orcidlink{0000-0003-4730-8830}} 
\newcommand{\orcidauthorN}{\orcidlink{0009-0005-9432-320X}} 

\def\BibTeX{{\rm B\kern-.05em{\sc i\kern-.025em b}\kern-.08em
    T\kern-.1667em\lower.7ex\hbox{E}\kern-.125emX}}
\begin{document}

\title{Evaluation of Sparse Acoustic Array Geometries for the Application in Indoor Localization}

\author{Georg~K.J.~Fischer\orcidauthorB, Niklas~Thiedecke\orcidauthorN, Thomas Schaechtle\orcidauthorC, Andrea~Gabbrielli\orcidauthorA, Fabian~Höflinger\orcidauthorH, Alexander~Stolz\orcidauthorM, and Stefan\,J.~Rupitsch\orcidauthorL, \IEEEmembership{Member, IEEE}
\thanks{This work was partially supported by the German Ministry of Education and Research (BMBF) under the grant FKZ: 16ME0028 Verbundprojekt ISA4.0, as well as by the German Ministry of Economic Affairs and Climate Action (BMWK) under grant FKZ: 03EE3066D Verbundvorhaben LoCA.}
\thanks{Georg~K.J.~Fischer, Thomas Schaechtle, Fabian~Höflinger and Alexander~Stolz are with the Fraunhofer Institute for Highspeed Dynamics, Ernst-Mach-Institute (EMI), Freiburg, Germany, E-Mail: georg.fischer@emi.fraunhofer.de.}
\thanks{Niklas~Thiedecke is with Bosch Sensortec GmbH, Reutlingen, Germany.}
\thanks{Andrea~Gabbrielli, Thomas~Schaechtle, Fabian~Höflinger and Stefan\,J.~Rupitsch are with the Department of Microsystems Engineering (IMTEK), University of Freiburg, Germany.}}

\maketitle

\begin{abstract}
\ac{AoA} estimation technology, with its potential advantages, emerges as an intriguing choice for indoor localization. Notably, it holds the promise of reducing installation costs. In contrast to \ac{ToF}/\ac{TDoA} based systems, \ac{AoA}-based approaches require a reduced number of nodes for effective localization. This characteristic establishes a trade-off between installation costs and the complexity of hardware and software. Moreover, the appeal of acoustic localization is further heightened by its capacity to provide cost-effective hardware solutions while maintaining a high degree of accuracy. Consequently, acoustic \ac{AoA} estimation technology stands out as a feasible and compelling option in the field of indoor localization.
Sparse arrays additionally have the ability to estimate the \ac{DoA} of more sources than available sensors by placing sensors in a specific geometry. 
In this contribution, we introduce a measurement platform designed to evaluate various sparse array geometries experimentally. The acoustic microphone array comprises 64 microphones arranged in an 8x8 grid, following an Uniform Rectangular Array (URA) configuration, with a grid spacing of 8.255 mm. This configuration achieves a spatial Nyquist frequency of approximately 20.8 kHz in the acoustic domain at room temperature. Notably, the array exhibits a mean spherical error of 1.26° when excluding higher elevation angles. The platform allows for masking sensors to simulate sparse array configurations. We assess four array geometries through simulations and experimental data, identifying the Open-Box and Nested array geometries as robust candidates. Additionally, we demonstrate the array's capability to concurrently estimate the directions of three emitting sources using experimental data, employing waveforms consisting of orthogonal codes.
\end{abstract}

\begin{acronym}[JSONP]\itemsep0pt
    \acro{DoA}{Direction-of-Arrival}
    \acro{ToF}{Time-of-Flight}
    \acro{DoF}{Degrees-of-Freedom}
    \acro{SBL}{Sparse Bayesian learning}
    \acro{RF}{Radio Frequency}
    \acro{MRA}{Minimum Redundancy Array}
    \acro{CRLB}{Cramér-Rao Lower Bound}
    \acro{GDoP}{Geometric Dilution of Precision}
    \acro{UWB}{Ultra-Wideband}
    \acro{AoA}{Angle-of-Arrival}
    \acro{TDoA}{Time-Difference-of-Arrival}
    \acro{ULA}{Uniform Linear Array}
    \acro{URA}{Uniform Rectangular Array}
    \acro{UCA}{Uniform Circular Array}
    \acro{SNR}{Signal-to-Noise Ratio}
    \acro{GCC}{Generalized Cross-Correlation}
    \acro{ESPRIT}{Estimation of Signal Parameters via Rotational Invariance Techniques}
    \acro{MUSIC}{Multiple Signal Classification}
    \acro{LS}{Least Squares}
    \acro{TLS}{Total Least Squares}
    \acro{PCB}{Printed Circuit Board}
    \acro{PSD}{Power Density Spectrum}
    \acro{CDF}{Cumulative Distribution Function}
    \acro{MEMS}{Micro-Electro-Mechanical Systems}
    \acro{TDM}{Time-Division Multiplexing}
    \acro{FPGA}{Field Programmable Gate Array}
    \acro{HPS}{Hard Processor System}
    \acro{RAM}{Random Access Memory}
    \acro{RMSE}{Root Mean Square Error}
    \acro{N-RMSE}{Normalized Root Mean Square Error}
    \acro{GUI}{Graphical User Interface}
    \acro{CPA}{Coprime Planar Array}
    \acro{ZC}{Zadoff-Chu}
    \acro{MLM}{Mainlobe Magnitude}
    \acro{MLW}{Mainlobe Width}
    \acro{MSLR}{Mainlobe / Sidelobe Ratio}
    \acro{MSLS}{Mainlobe / Sidelobe Separation}
    \acro{IQ}{In-Phase Quadrature}
    \acro{SPL}{Sound Pressure Level}
    \acro{MSE}{Mean Squared Error}
    \acro{GT}{Ground Truth}
    \acro{FoV}{Field of View}
    \acro{BPF}{Bandpass Filter}
    \acro{CI}{Confidence Interval}
\end{acronym}
\section{Introduction}
\IEEEPARstart{D}{irection-of-Arrival} (DoA) estimation proves to be a common challenge within the domain of indoor localization. By strategically navigating the trade-off between hardware and software complexity, it becomes feasible to minimize the requisite number of anchor nodes for localization tasks. Fundamentally, only two nodes are necessary to intersect lines in \mbox{3-D} space, thereby facilitating position estimation and reducing installation costs. In contrast, \ac{ToF} systems demand a minimum of four nodes for 3-D localization.
DoA-based systems entail a more intricate hardware setup, incorporating multiple arranged sensors such as antennas or microphones. This necessitates sophisticated hardware design, particularly in areas like Radio Frequency (RF) design, tuning, and data acquisition and processing, leading to larger devices. Furthermore, the interdependence of algorithms and hardware development is crucial, as data models heavily rely on parameters like array geometries, sampling specifications, and data handling limits.
Despite the hardware complexity, the reduction in required nodes significantly diminishes installation costs. Additionally, receiver synchronization is unnecessary since only angles are estimated.

The \ac{GDoP} in DoA-based systems exhibits a distinct structure compared to ToF/TDoA-based systems. While ToF/TDoA systems dilute precision outside the polygon of receivers, DoA systems experience precision dilution in the line of receivers and with distance to the receivers (forming a cone shape) \cite{Shin2002, Torrieri1984, Zhou2021}. This characteristic makes it an intriguing design choice for engineers to explore and implement.


The selection of technology spans from RF-based devices, such as WiFi, Bluetooth, 5G, mmWave, or even \ac{UWB}, to optical and acoustic systems and combinations thereof \cite{Fischer.2022,Girolami.2024,Sesyuk.2024}. A key distinguishing factor among these technologies is primarily the propagation speed of waves in the medium, which significantly influences the achievable performance bounds \cite{Mirkin.1991, Wen.2018b, Zhang.2023}.
Another critical characteristic is the maturity of the technologies. For instance, technologies like Bluetooth have specific sections in their standards dedicated to \ac{AoA} estimation \cite{Bluetooth2019}, while others may lack such provisions. Compatibility is also a noteworthy concern. Given that nearly all smartphones support technologies like Bluetooth, incorporating acoustic localization, despite having the required hardware, necessitates the installation of custom software on the device \cite{Hoflinger.2012}.

Acoustic localization technology, despite not yet being widely deployed, possesses the advantage of achieving high accuracy with straightforward and cost-effective hardware. Its positive traits, including resilience to \ac{RF} interference, make it a rational choice for deployment in mid-sized areas where both high accuracy and cost-effectiveness are crucial considerations.
Acoustic \ac{AoA} estimation has been subject to exploration in prior research. Conventional algorithms for \ac{DoA} estimation, particularly when dealing with multiple tags, often necessitate time-synchronized signal emissions. However, this requirement increases tag complexity. Alternatively, accepting a certain degree of message collisions is another approach, albeit making location updates less predictable \cite{Gabbrielli.2023b}.
Classical subspace estimators offer the capability to simultaneously estimate the DoA of multiple signals. However, the number of locatable sources using these methods is strictly limited by the number of available sensors \cite{Roy.1986, Schmidt.1986}.

In this study, our primary focus is to explore the potential for increasing the number of concurrently locatable tags by leveraging sparse arrays. To facilitate this investigation, we introduce an evaluation platform comprising 64 microphones arranged in an \ac{URA} configuration. This platform enables us to analyze and evaluate the performance of various sparse array geometries.

It is noteworthy that this paper extends our previously published work in \cite{Fischer.2023}. In addition to the previously published material, we have expanded the literature review and related work section to provide a broader overview. New data has been collected, including more azimuth and elevation angle pairs. With the complete angle space sampled, a full characterization of the evaluated geometries is now possible, revealing the characteristics of each. Furthermore, the multiple sources experiment has been extended to include another source, now encompassing three sources in total.


\section{Related Work}

The related work section is divided into two main components. First, we consider the field of acoustic \ac{AoA} indoor localization, exploring prior research and advancements. The second part provides a concise review of sparse arrays, elucidating their applications and contributions in diverse contexts.

\subsection{Acoustic \ac{AoA} for Indoor Localization}
Various acoustic localization systems grounded in \ac{AoA} localization have been proposed, featuring array geometries encompassing microphone pairs, triangular configurations, \acp{ULA}, and \acp{UCA}. 
Saad\etal \cite{Saad.2012} integrate the \ac{AoA} methodology with \ac{ToF} measurements, achieving a 95th percentile localization error of less than \SI{10}{\centi\meter} through their triangular array configuration. In another endeavor \cite{Li.2017}, the authors leverage a uniform linear array with four microphones, yielding a 95-th percentile localization error under \SI{65}{\centi\meter}. The work in \cite{Ogiso.2019} explores self-positioning through a microphone pair and strategically deployed acoustic beacons. Gabbrielli\etal \cite{Gabbrielli.2023,Gabbrielli.2023b,Gabbrielli.2021} propose a five-microphone \ac{UCA} configuration, achieving a 95\% error rate of under \SI{17}{\centi\meter} in 3-D absolute localization accuracy.
Generally, a limited number ($<$10) of microphones is employed to facilitate \ac{AoA} estimation on embedded devices. However, the development of an indoor localization system capable of concurrently performing \ac{AoA} estimation for multiple emitting sound sources (tags) remains a pending objective in the current state of research.


\subsection{Sparse Arrays and their Applications}
Sparse arrays have been a subject of research for several decades \cite{Aboumahmoud.2021}. Various geometries have been explored, including Nested arrays \cite{Pal.2010, Pal.2012b, Pal.2012, Nannuru.2018}, Coprime arrays \cite{Vaidyanathan.2011, Liu.2017b, Wu.2016}, Hourglass arrays \cite{Liu.2017}, and Thermos arrays \cite{Sun.2018}. In the \ac{RF} domain, sparse arrays offer the potential to mitigate mutual coupling between antennas \cite{Sun.2018}. Theoretical performance bounds, such as the \ac{CRLB}, have been established for various geometries, including Coprime arrays, Nested arrays, and \acp{MRA} \cite{Liu.2017c}. Furthermore, sparse arrays have found application in other domains, such as medical ultrasound \cite{Ramalli.2022b}, showcasing the potential to reduce costs, albeit with a trade-off involving diminished image quality.


Within the field of acoustic beamforming applications,  sparse arrays have gained prominence, demonstrating adaptability across diverse scenarios. In the domain of speech localization, the work in \cite{Zhao.2019} reports the observation that Semi-Coprime microphone arrays exhibit superior beam patterns and reduced side lobe levels compared to coprime arrays.
Bush\etal\cite{Bush.2015} conducted an assessment and validation of simulation results for Coprime linear microphone arrays with wideband sources. In a related investigation \cite{Bush.2018}, source enumeration utilizing a Bayesian approach is explored, employing Coprime linear microphone arrays.
A technique utilizing sparse linear arrays for uncorrelated wideband sources is introduced in \cite{Wang.2021}. This approach integrates data from different frequency bins into a unified matrix, employing atomic norm minimization for precise \ac{DoA} estimation. The assessment of this technique involved the utilization of a two-level Nested array.
In the experimental domain, the work in \cite{Nannuru.2021,Nannuru.2018} conducted a thorough evaluation of planar sparse arrays employing the \ac{SBL} approach. This array, consisting of 63 microphones, operated in the lower frequency range (under \SI{3}{\kilo\hertz}). The authors could show, that the \ac{SBL} algorithm is able to resolve an equal number of sources as available sensors in the 2-D case.


Two-Dimensional \ac{DoA} estimation is a consequential extension to the 1-D arrays. In light of this, Aboumahmoud \etal \cite{Aboumahmoud.2021} offer a comprehensive review specifically focused on various sparse geometries for 2-D estimation. Within their work, the authors conduct a comparative evaluation considering critical factors such as the number of sensors, their maximum \acp{DoF}, and aperture sizes. Notably, the predominant nature of these investigations is analytical and performed \textit{in silico}. Consequently, the existing body of research is characterized by a limited number of experimental validations and demonstrated applications.

The principal contributions of this work encompass:
\begin{itemize}
\item The conceptualization of a compact 64-microphone array system tailored for indoor localization applications in the inaudible frequency range.
\item Investigation into the system's performance, conducted through a comprehensive analysis involving both simulation and experimentation.
\item Evaluation of various sparse array geometries utilizing empirical data.
\item A practical demonstration showcasing the system's proficiency in estimating the \ac{DoA} for multiple concurrently emitting sound sources.
\end{itemize}

\section{Signal Model}\label{sec:signal-model}
\begin{figure}
	\begin{centering}
		\includegraphics[width=\linewidth]{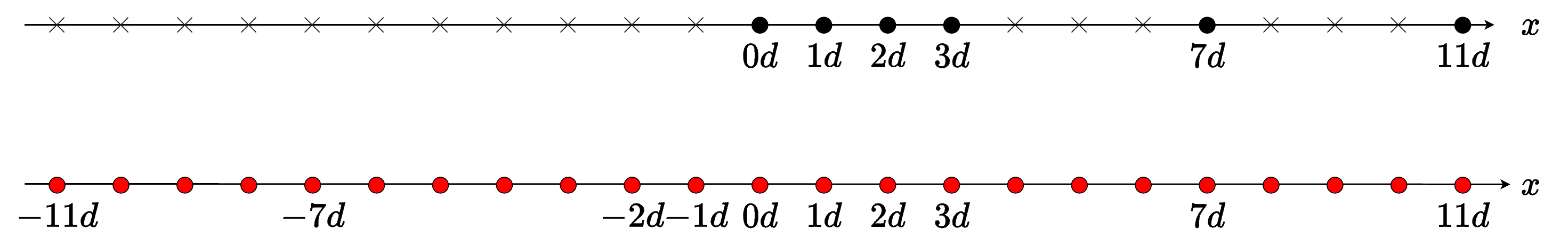}
		\caption{1-D nested array (top) with aperture $N = 11d$ and its difference co-array (bottom). Circles mark sensors and crosses mark gaps. The positions of all virtual sensors are calculated as the differences of positions of the physical sensors.}
		\label{fig:nested_1d_diff}
	\end{centering}
\end{figure}

The central idea behind sparse microphone arrays lies in exploiting redundancy within data recorded using non-sparse arrays, e.g. \acp{URA}, by reconstructing the information of missing microphones using a combination of two or more other microphones. 
\paragraph{Difference co-array}
Sparse signal processing uses a difference co-array to reconstruct this information, which is defined as the set of differences in microphone positions \cite{Aboumahmoud.2021}: 
\begin{align}
	\mathbb{D} = \{m\ \vert\ m = \kappa_i - \kappa_j,\quad \forall\ \kappa_i, \kappa_j \in \mathbb{S}\},
\end{align}
where $\mathbb{S}$ is the set of sensor positions in the array. The set of resulting positions $\mathbb{D}$ describes the position of all virtual sensors in the difference co-array. \autoref{fig:nested_1d_diff} shows an example of a 1-D nested array and its difference co-array.
In order to calculate the virtual sensor data of the difference co-array of a given physical microphone array, the sample auto correlation matrix of the array's sensor data has to be vectorized:
\begin{align}
	\vec{z} = \mathrm{vec}(\mat{\hat{R}}_y) &= (\mat{A}^*\odot \mat{A})\vec{p} + \sigma_n^2 \mat{1}_n
	\intertext{with}
	\mat{\hat{R}}_{y} &= \frac{1}{K} \sum_{k=1}^{K}\vec{y}[k]\vec{y}[k]^\mathrm{H},\\
	\vec{p} &= \left[
	\begin{array}{cccc}
		\sigma_1^2 &\sigma_2^2 &\cdots &\sigma_M^2
	\end{array}
	\right]^\mathrm{T}.
\end{align}
Here, $\vec{y}[k]$ denotes the vector of microphone samples of the $k$-th microphone and $\vec{p}$ holds the powers of the $M$ source signals, which are the diagonal entries of the source auto correlation matrix $\mat{R}_s$. The noise term is represented by $\sigma_n^2{\vec{1}}_n$ and the Hermitian transpose of $\vec{y}$ is denoted by $\vec{y}^\mathrm{H}$. The effective array manifold matrix $\mat{A}_{\mathrm{eff}} = \mat{A}^*\odot \mat{A}$ can be interpreted as the steering matrix corresponding to the whole difference co-array and is therefore able to exploit all \ac{DoF} of the physical array \cite{Pal2010}. The $\odot$-operator denotes the Khatri-Rao product. \par
\paragraph{Spatial Smoothing} \label{sec:spat_smooth}
The virtual sensor data vector $\vec{z}$ cannot be used for \ac{DoA}-estimation directly because it behaves like fully coherent sources, which violates the assumptions required by the subspace methods. This problem can be taken care of in different ways, one of them being the spatial smoothing technique as discussed by Pal and Vaidyanathan \cite{Pal2010}, which can be applied without modification for any difference co-array that has a coherent \ac{ULA} segment in the 1-D case or a coherent \ac{URA} segment in the 2-D case.\par
In order to be able to perform spatial smoothing, the data corresponding to repeating virtual sensors in the difference co-array has to be removed and afterwards the data is re-ordered. This is equivalent to only keeping unique entries of the observation vector $\vec{z}$ and sorting them, which leads to a new vector $\vec{z}_1$. The sorting is done such, that the data in ${\vec{z}_1}_j$ corresponds to the $j$-th element in the center \ac{ULA}-segment of the difference co-array. 
\begin{figure}
	\begin{centering}
		\includegraphics[width=\linewidth]{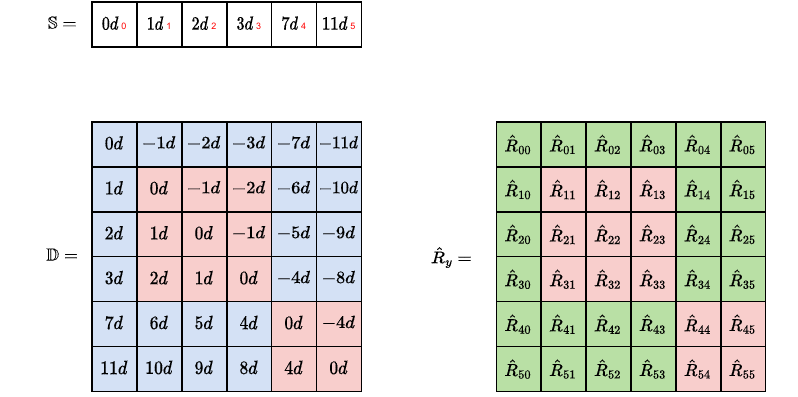}
		\caption{Difference co-array and entries of auto correlation matrix of 1-D nested array. Green squares denote virtual positions which appeared for the first time, red squares denote positions which are repetitions of the green positions.}
		\label{fig:construction_z}
	\end{centering}
\end{figure}

\autoref{fig:construction_z} shows entries of the auto correlation matrix in the example of a 1-D Nested array with physical positions $0d$, $1d$, $2d$, $3d$, $7d$, $11d$. In this example, red boxes denote repeated (redundant) virtual sensor positions in the difference co-array. Here, $\mathrm{vec}(\mathbb{D}')$ denotes the vector containing all elements of the set $\mathbb{D}$ in increasing order regarding virtual sensor positions. The entries of the auto correlation matrix $\mat{\hat{R}}_y$, which correspond to redundant virtual sensor positions, are removed.\par In the 2-D case, the ordering is performed such that the center \ac{URA}-segment ordering is matched \cite{Pal2010}.

\begin{figure*}[ht]
	\begin{centering}
		\includegraphics[width=\linewidth]{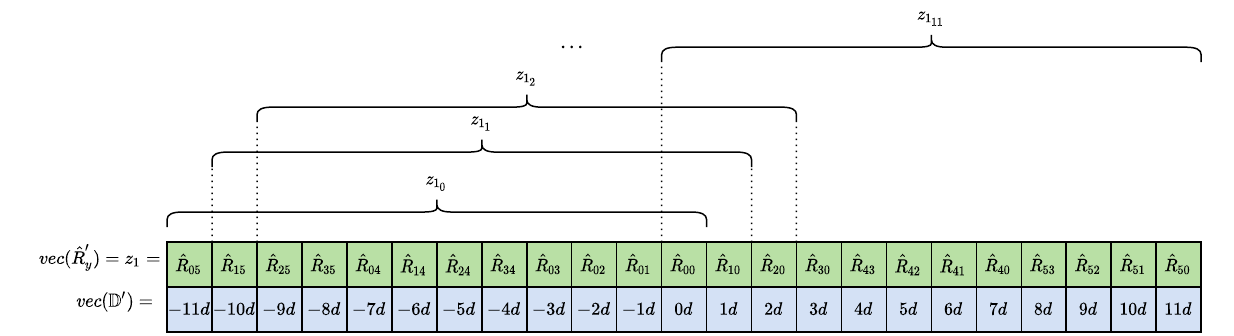}
		\caption{Construction of spatial smoothing sub-arrays through the example of 1-D Nested array. Each sub-array consists of 12 virtual sensor positions and therefore 12 auto correlation components. The overlap between neighboring sub-arrays is chosen to be 10, maximizing the number of smoothed components.}
		\label{fig:construction_subarrays_z}
	\end{centering}
\end{figure*}

The spatial smoothing step now divides the co-array into overlapping sub-arrays, and calculates an auto correlation matrix on the corresponding segment of the virtual sensor observation vector $\mat{z}_1$. Each individual auto correlation matrix is computed as 
\begin{align}
	\mat{R}_i = \vec{z}_{1_i}\vec{z}_{1_i}^\mathrm{H}
\end{align}
The selection of sub-arrays and the corresponding segments of $\vec{z}_1$ is visualized in \autoref{fig:construction_subarrays_z}.
Finally, the spatial smoothing output auto correlation matrix $\mat{R}_{ss}$ is computed by averaging over all $\mat{R}_l$ with
\begin{align}
	\mat{R}_{ss} = \frac{1}{L} \sum_{l=1}^{L} \mat{R}_l
    \label{eq:rss}
\end{align}
where $L$ is the number of sub-arrays of the difference co-array, which is a design parameter depending on the array structure and size.

\paragraph{2-D \ac{DoA} estimation on sparse arrays}
When spatial smoothing was performed, \ac{MUSIC} and \ac{ESPRIT}\cite{Roy.1986,Schmidt.1986} can be used on the new auto correlation matrix $\mat{R}_{ss}$ (see \autoref{eq:rss}), as long as the sparse array is designed such, that the difference co-array has a coherent \ac{URA} center piece. \par 
\ac{MUSIC} has to be used with the steering matrix $\mat{A}$ corresponding to a \ac{URA} with the size of the segment that was used for the spatial smoothing step in the previous section. If \ac{MUSIC} is then performed on the spatially smoothed auto correlation matrix $\mat{R}_{ss}$, the \ac{DoA} of the signal impinging on the original physical array can be estimated.

\section{Hardware}

One key property, which most sparse arrays have in common, is that all sensors are placed on an equidistantly spaced grid, where each node has a distance $d$ to its neighboring nodes. Exploiting this property, each node can be filled out by actual sensors, such, that multiple sparse geometries can be mapped onto this grid by not using the data of specific nodes for estimation, given these nodes are not part of the sparse array currently examined.
\begin{figure}
	\begin{centering}
		\includegraphics[width=\linewidth]{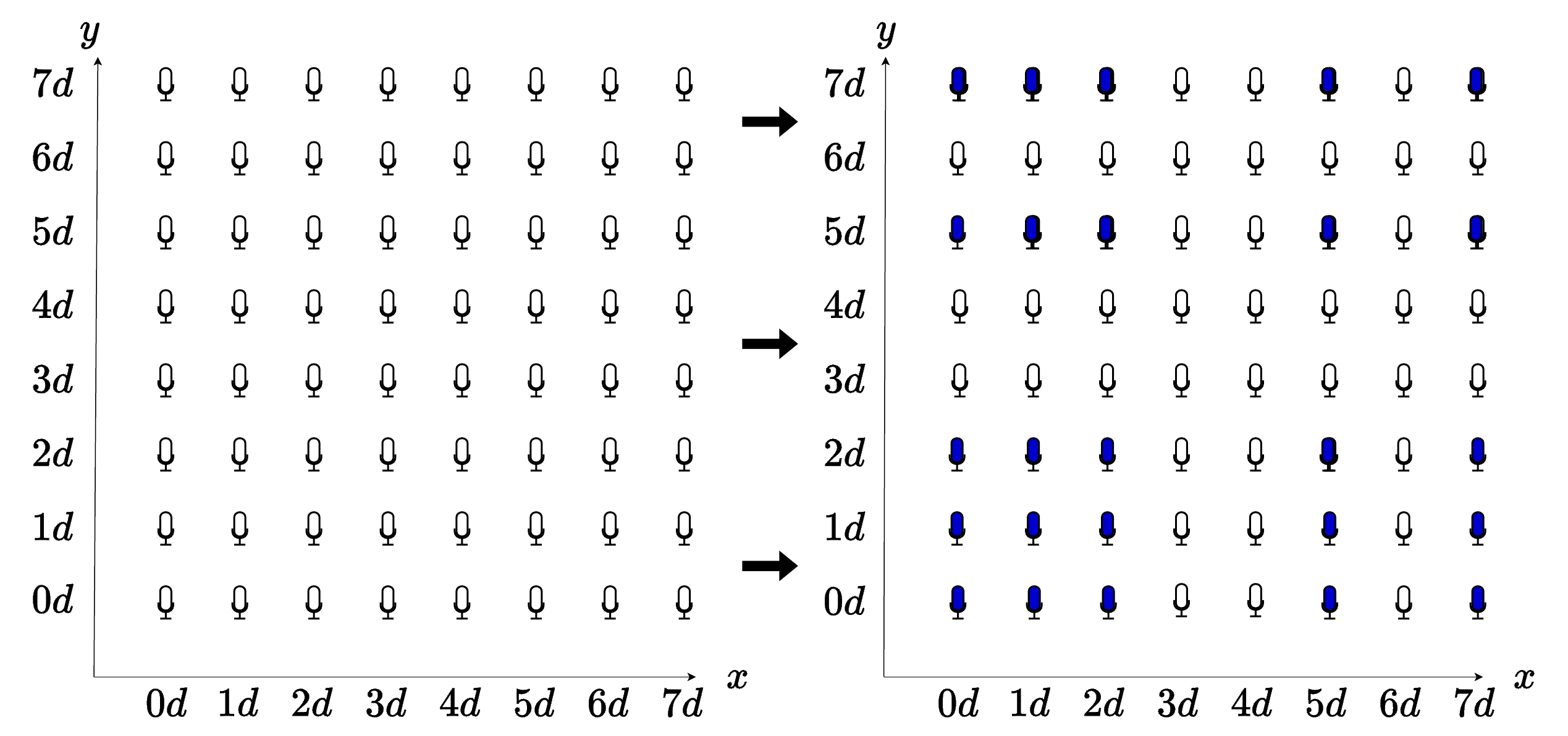}
		\caption{Hardware concept of sparse microphone arrays: The left side shows the actual physical sensors present. On the right side, blue microphones are actually used for estimation, in order to get the data corresponding to a 2-D Nested array.}
		\label{fig:hw_concept_array}
	\end{centering}
\end{figure}
\autoref{fig:hw_concept_array} illustrates this concept. It can be seen, that the only parameters limiting the number of depictable geometries are the apertures (i.e., the number of nodes) in x- and y-direction, respectively. Thus, these parameters were chosen as large as possible, while still being able to sample with more than double the frequency of the target sound source frequency band of around \SI{20}{\kilo\hertz}. The resulting array design is built as an 8x8 grid, which results in a total of 64 microphones.\par
The main challenges of designing such a system for the acoustical use-case include:
\begin{itemize}
	\item Synchronization of exact analog sampling times
	\item Data integrity with several high frequency signals in the MHz range close to each other
	\item Ability to process and store data of all 64 microphones concurrently
\end{itemize}
In this contribution, the \ac{MEMS} microphone \textit{ICS52000} was chosen, as it is specifically designed for array applications. The microphone uses a digital \ac{TDM} interface, which enables having up to 16 microphones use a single data line and synchronize with each other by sharing one clock line and daisy-chaining a handshake signal through the whole chain of sensors. \\\\
In order to communicate to all microphones at once and store all the data in parallel at sufficient speed, the communication unit for the system was chosen to be a \ac{FPGA}, since the \ac{TDM} interface, as well as an on-chip \ac{RAM} can be implemented in a parallelized fashion.\par

\begin{figure}[h]
	\begin{centering}
		\includegraphics[width=\linewidth]{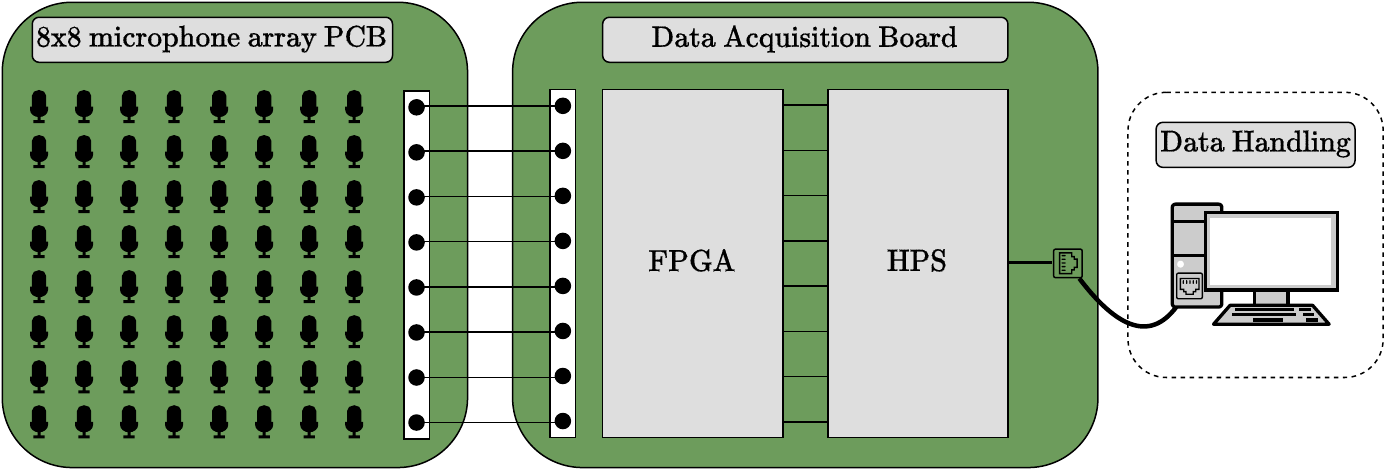}
		\caption{Complete hardware system concept: The data flow moves from left to right. Microphone data is recorded on the \ac{PCB}, read by the Data Acquisition Board, and sent to the computer for data handling and analysis.}
		\label{fig:hw_concept_complete}
	\end{centering}
\end{figure}
The complete system concept is depicted in \autoref{fig:hw_concept_complete}. For acquiring data, the \textit{DE0-Nano-SoC} evaluation board was chosen, which contains both a programmable \textit{Cyclone V} FPGA and a software programmable \ac{HPS}, facilitating low-level implementation of the \ac{TDM} interface, data storage and transfer to a computer.

\subsection{System Implementation} \label{sec:PCB}
\begin{figure}
	\begin{centering}
        \resizebox{\linewidth}{!}{\input{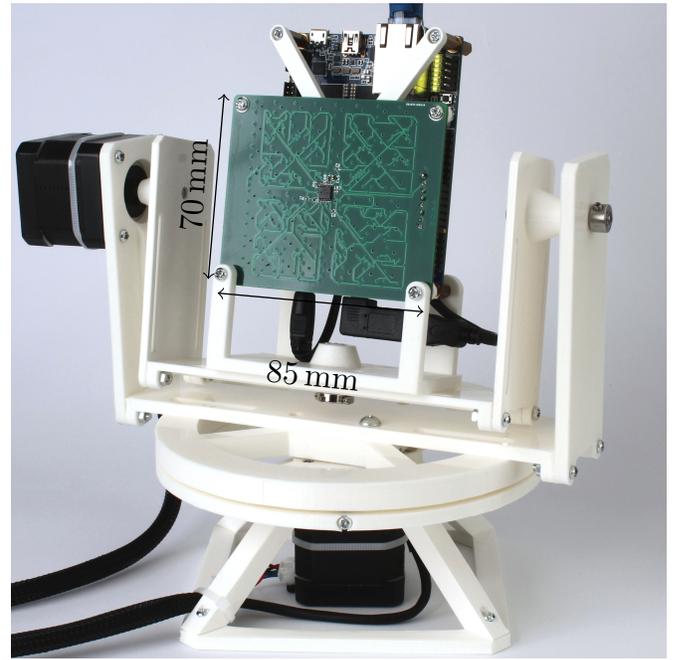}}
		\caption{The photo showcases the microphone array mounted on a two-axis rotation table\protect\footnotemark, essential for conducting accuracy measurements. Microphone capsules are situated on the backside of the \ac{PCB}, linked to the front side via dedicated holes. The PCB is sandwiched onto the Cyclone V FPGA board, which includes an Ethernet connection for data transfer and control.}
		\label{fig:pcb}
	\end{centering}
\end{figure}

Because the maximum number of interconnectable \textit{ICS52000} microphones is 16, the whole eight by eight array is split into four sub-arrays, each taking up a quarter of the full array. Each sub-array receives a copy of the system clock, which is provided by a clock-buffer to ensure clean edges. In order to synchronize the four sub-arrays among each other, the first microphone in each of them obtains a global word select signal, which is directly connected to the \ac{FPGA}. To avoid timing differences induced by length differences of \ac{PCB} traces, all clock lines leading to a single microphone are matched to the same length.
The geometric distance $d$ between each neighboring microphone is set to \SI{8.255}{\milli\meter}. This distance corresponds to half the smallest wavelength $\lambda_{min}$ of a signal, which can still be sampled in the half space. In other words, the largest allowed frequency of sound sources is determined as $f_{\mathrm{max}} =  c\lambda_{\mathrm{min}}^{-1} = c(2d)^{-1} \approx \SI{20.788}{\kilo\hertz}$ in free air at room temperature. If the frequency band of a sound source is larger than this limit, robust estimation of its location will not be possible. \autoref{fig:pcb} displays a picture of the developed array.

\footnotetext{Design files of the rotation table are available under: \url{https://github.com/e1kable/two-axis-rotation-table}}

\subsection{Signal Processing} \label{sec:signal-processing}
\begin{figure}
	\begin{centering}
		\resizebox{\linewidth}{!}{\usetikzlibrary{shapes,arrows,chains,arrows.meta,positioning}
\begin{tikzpicture}[%
    >=triangle 60,              
    start chain,    
    node distance=10mm and 30mm, 
    every join/.style={norm},   
    ]
    \tikzset{
      base/.style={draw, on chain, on grid, align=center, minimum height=4ex},
      proc/.style={base, rectangle, text width=7em},
      test/.style={base, diamond, aspect=2, text width=5em},
      term/.style={proc, rounded corners},
      coord/.style={coordinate, on chain, on grid, node distance=6mm and 25mm},
      nmark/.style={draw, cyan, circle, font={\sffamily\bfseries}},
      norm/.style={->, draw},
      free/.style={->, draw},
      cong/.style={->, draw},
      it/.style={font={\small\itshape}}
    }
    \node [proc, densely dotted, it, on chain] (p0) {$\mathbf{S} \in \mathbb{R}^{K\times N}$};
    \node [proc, join, on chain]  (L)    {BPF};
    \node [proc, join, on chain]      {$\mathbb{H} \{\cdot\}$};
    \node [proc, join, on chain=going below] (p1) {$\mathbf{C}\circ\mathbf{S}\in\mathbb{C}^{K\times N}$};
    \node [proc, join, on chain=going left] (p2) {Estimator};
    \node [term, join, on chain=going left] (t1) {DoA $(\hat{\varphi}, \hat{\theta})$};

\end{tikzpicture}}
		\caption{Detailed Overview of the Signal Processing Chain: This figure Depicts the progression from \ac{BPF} and Hilbert transform application, through calibration, to the final \ac{DoA} estimation.}
		\label{fig:signal-processing-chain}
	\end{centering}
\end{figure}
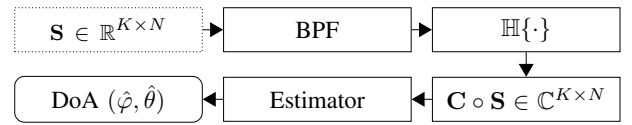

The signal processing chain is structured as follows (cf. \autoref{fig:signal-processing-chain}): First, the $K\times N$ sample matrix, denoted as $\Vec{S}$, is obtained by the array, where $K$ represents the number of channels and $N$ represents the number of samples. Next, each channel undergoes individual \ac{BPF} to a specific frequency using a 10th order Butterworth filter with a bandwidth $B$. Subsequently, we apply a Hilbert transform to each channel to derive complex \ac{IQ} samples. Following this, a calibration matrix $\Vec{C}$, similar in size to the complex sample matrix, is applied to ensure accuracy and consistency in the measurements, where $\circ$ denotes the Hadamard product. Finally, the calibrated, complex sample matrix is passed through to be further processed by the specific \ac{DoA} estimators, yielding pairs of azimuth and elevation angles, as detailed in \autoref{sec:signal-model}.

\section{Simulation}

In this section, we introduce various array geometries and simulatively assess their performance. Each geometry's beampattern is evaluated along with quantitative metrics. The simulations are conducted using the intended design parameters of the array, including dimensions and sampling frequency. The probing signal employed is a complex sinusoid at \SI{20}{\kilo\hertz}. In addition to specific design rules such as Billboard or co-prime, a randomly selected set of sensors is included as a last geometry, comprising approximately the same number of sensors as the other sparse geometries.

\begin{figure*}
	\centering
    \if\isPrint1
        \includegraphics[width=\linewidth]{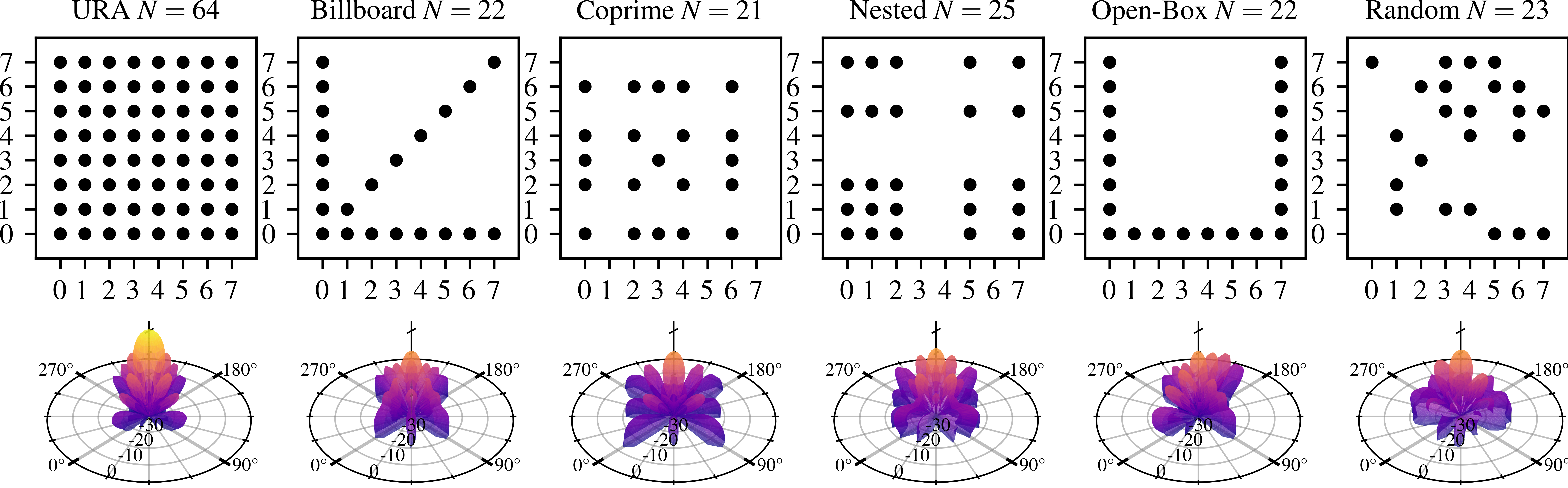}
    \else
        \includesvg[width=\linewidth]{figures/simulation/composite-Npoints-50.svg}    
    \fi
    \caption{The figure depicts the array geometries under investigation alongside their corresponding beampatterns as achieved with a classic delay-and-sum beamformer. The magnitude of the beampattern is represented in dB and mapped to the radius.}
    \label{fig:array-geometries-beampattern}	
\end{figure*}

\autoref{fig:array-geometries-beampattern} illustrates these geometries alongside their directional response to an impinging wavefront in the broadside direction. The response is calculated using a traditional delay-and-sum beamformer, providing insights into each geometry's performance, such as potential angle confusions due to large sidelobes.

Table \ref{tab:array-geometry-metrics} provides quantitative metrics for evaluating the beampatterns of various array geometries. Asymmetric geometries, such as the Open-box array, naturally result in asymmetric beampatterns. The \ac{URA} demonstrates the highest \ac{MLM}, outperforming the Nested array which shows a reduction of approximately \SI{-8}{\decibel}. The Open-Box array is distinguished by having the narrowest \ac{MLW}, as defined by the \SI{-3}{\decibel} attenuation threshold. Furthermore, the \ac{URA} secures the highest \ac{MSLR}, with the Billboard array following directly behind. The Coprime array is noted for achieving the greatest \ac{MSLS}. The correlation between an array's symmetry and its beampattern's symmetry is evident when contrasting the \ac{URA} with the Open-Box array in \autoref{fig:array-geometries-beampattern}.

\begin{table}
    \centering
    \renewcommand{\arraystretch}{1.2}%
    \begin{tabular}{@{}lcccc@{}}\toprule
         Array & \thead{MLM in dB} & MLW in $\deg$ & MSLR in dB & MSLS in $\deg$  \\
         \midrule
        URA & \textbf{0.00}* & 13.57 & \textbf{12.80} & 21.71  \\
        Billboard & -9.28 & 12.66 & 7.38 & 20.35  \\
        Coprime & -9.68 & 14.47 & 3.18 & \textbf{90.00}  \\
        Nested & -8.16 & 11.76 & 5.52 & 20.35  \\
        Open-Box & -9.28 & \textbf{9.95} & 3.26 & 17.64  \\
        Random & -8.89 & 12.66 & 5.56 & 23.07  \\
        \bottomrule
    \end{tabular}
    \caption{Comparison of serveral array metrics (* denotes the reference). In this comparison, the mainlobe of the \ac{URA} serves as the reference point for the Main Lobe Magnitude (MLW). Furthermore, Mainlobe Width (MLW), Mainlobe to Sidelobe Ratio (MSLR) and Mainlobe to Sidelobe Separation (MSLS) are detailed. }
    \label{tab:array-geometry-metrics}
\end{table}

The overall angular error $e$ is assessed using the spherical angular distance between the unit vectors of the estimated point $\Vec{\hat{p}_i}$ and the ground truth value $\Vec{p_i}$ defined as
\begin{equation}    
    e = \cos^{-1}\left(\Vec{\hat{p}_i}^\mathrm{T}\Vec{p_i}\right). 
    \label{eq:spherical-error}
\end{equation}
This metric effectively addresses situations where uncertainties in the azimuth angle may not lead to significant deviations, particularly noticeable in low elevation angle regimes.
\autoref{fig:array-geometries-angular-error} illustrates the achievable Mean Angular Error under additive Gaussian noise with a specific modeled \ac{SNR}. In this simulation, the ordering is solely correlated with the number of available sensors, with the URA configuration performing the best.
A similar outcome can be attained by assessing the \ac{CRLB}\cite{Stoica.1990b, Stoica.1989}, which defines the achievable \ac{MSE} for both azimuth and elevation angles separately.
The advantage of this simulation lies in its ability to account for irrelevant errors in fringe regions through a composite metric.

\begin{figure}
	\centering
    \input{figures/simulation/angular-error-over-snr-Niter-2000-deg.pgf}
    \caption{Simulation across the angle space $\varphi\in\left[\SI{0}{\degree},\SI{360}{\degree}\right]$ and $\theta\in\left[\SI{0}{\degree},\SI{90}{\degree}\right]$ with $K=1000$ samples. The error metric is the spherical angles between estimation and the \ac{GT} point.}
    \label{fig:array-geometries-angular-error}	
\end{figure}
\section{Experiments}
\label{chap:experiments}
In this section, the results of several experiments are presented. The measurement setup is first introduced, followed by a discussion of the calibration results. Next, various algorithms are compared against each other, along with an evaluation of their performance using different sparse geometries. The section concludes with an assessment of the system's capability to handle multiple concurrently emitting sources, accompanied by a discussion of the obtained results.

\subsection{Measurement Setup}
The experiments are conducted within an anechoic chamber from the company Wendt-Noise Control, providing an environment free from sound reflections and external interference. The array is assembled onto a two-axis rotation table, as depicted in \autoref{fig:pcb}. This rotation table allows rotation in both azimuth and elevation directions. A speaker is positioned approximately \SI{3.7}{\meter} away from the array to emit sound signals for localization purposes. Additionally, a linear reference microphone is employed to provide a reference for the measured amplitudes.

\subsection{Calibration and Analysis} \label{sec:Calib}
The calibration process comprises both phase calibration and amplitude adjustment. The amplitude adjustment serves two purposes: firstly, to standardize the microphones within the array relative to each other, and secondly, to align the microphone data with \acp{SPL} $L_p$, in decibels (dB) relative to \SI{20}{\micro\pascal}.

\begin{figure}
	\centering
     \if\isPrint1
        \includegraphics[width=\linewidth]{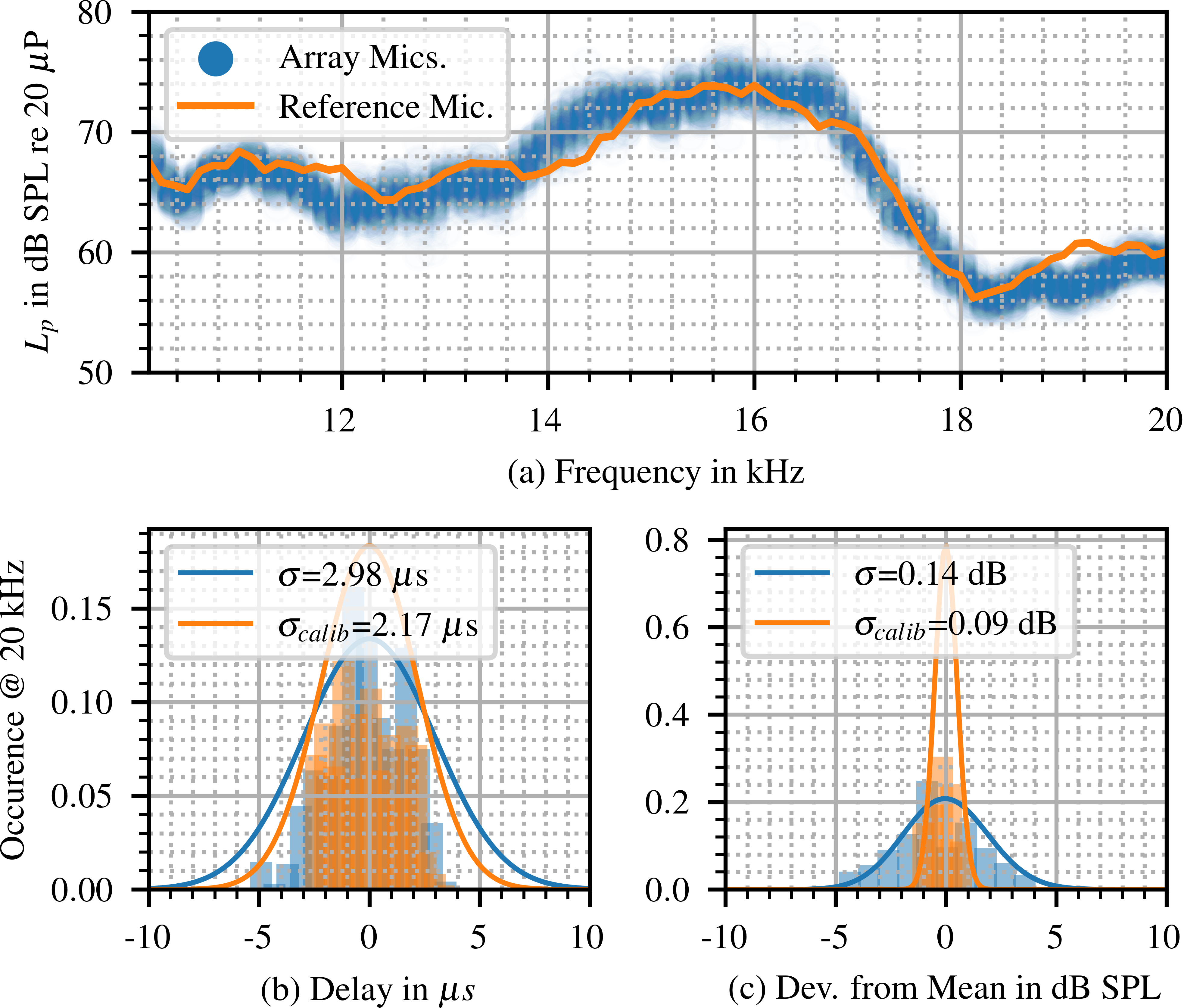}
    \else
        \input{figures/calibration/frequency-calib-composite-plot-ieee.pgf}
    \fi
    \caption{Calibration analysis: (a) depicts the measured and calibrated sound pressure levels obtained from the measurement microphone, along with those recorded by the individual array microphones.
    (b) and (c) showcase histograms illustrating the results of the phase and amplitude calibration specifically at \SI{20}{\kilo\hertz}. These plots include both the raw data distribution and the fitted normal distributions (for phase) and Log-normal distributions (for sound pressure).}
    \label{fig:calibration}	
\end{figure}

To obtain dB \ac{SPL} calibration values, a Behringer ECM8000 linear measurement microphone is referenced with a B\&K Type-4231 calibrator at \SI{94}{\decibel~SPL}. Subsequently, the frequencies are swept through in the range of \SIrange{10}{20}{\kilo\hertz} and measured by both the array and the measurement microphone, as depicted in \autoref{fig:calibration} (a). The variation in sound pressure level arises due to the characteristics of the employed Beyma T2030 speaker.

In \autoref{fig:calibration} (b), both uncalibrated and calibrated phase measurements are displayed, indicating marginal differences at the shown frequency of \SI{20}{\kilo\hertz}. The sound pressure in dB is modeled with a Log-normal distribution, with its fit shown in (c). In contrast to the phase calibration, the total variance of the measured levels may be reduced by around 40 percent.

It is important to note the significance of amplitude calibration in achieving accurate Direction-of-Arrival (DoA) measurements, given the assumption of equal gains across all sensors in the observation matrix $\Vec{A}$. The entries of the calibration matrix are chosen accordingly
\begin{equation}
    \Vec{C}_{i,j} = A_{i,c} \mathrm{e}^{-\mathrm{j} \Psi_{i,c}}
\end{equation}
where $A_{i,c}$ represents the amplitude calibration factor and $\Psi_{i,c}$ denotes the measured phase factor. The index variable $i \in \{1,\dots,K\}$ corresponds to the channel index and $j \in \{1,\dots,N\}$ to the sample index.
The calibration process yields slightly more accurate estimations, particularly noticeable in lower elevation angle regimes. For instance, below \SI{70}{\degree} at \SI{20}{\kilo\hertz}, the measured mean spherical error can be reduced from \SI{1.24}{\degree} to \SI{1.20}{\degree} .

\subsection{DoA-Estimation (Single Source)}

In order to show that the developed sensor system is capable of collecting data which are suitable of estimating the \ac{AoA} of a sound source, a set spanning the complete range of azimuth and elevation angles is composed. The incident sound pressure $L_p$ at the array is set to \SI{60}{\decibel~SPL}, which results in a \ac{SNR} of around \SI{30}{\decibel} (inband), in the given anechoic environment. For each angle pair, the microphone array captures audio samples which are then sent to a computer for analysis and ultimately estimating the \ac{AoA} of each source position. To ensure statistical convergence of the results, each measurement has a duration of $T=\SI{2}{\second}$, resulting in $N=96000$ samples. With only one source involved in this experiment, inter-source correlations pose no issue. Consequently, a simple sinusoidal signal is broadcast from the source without the need for additional processing or coding. The narrowband requirement of subspace algorithms is met by applying a band-pass filter to the recorded microphone data as detailed in \autoref{sec:signal-processing}.

\subsubsection{Comparison of algorithms}
\begin{figure}
	\centering
     \if\isPrint1
        \includegraphics[width=\linewidth]{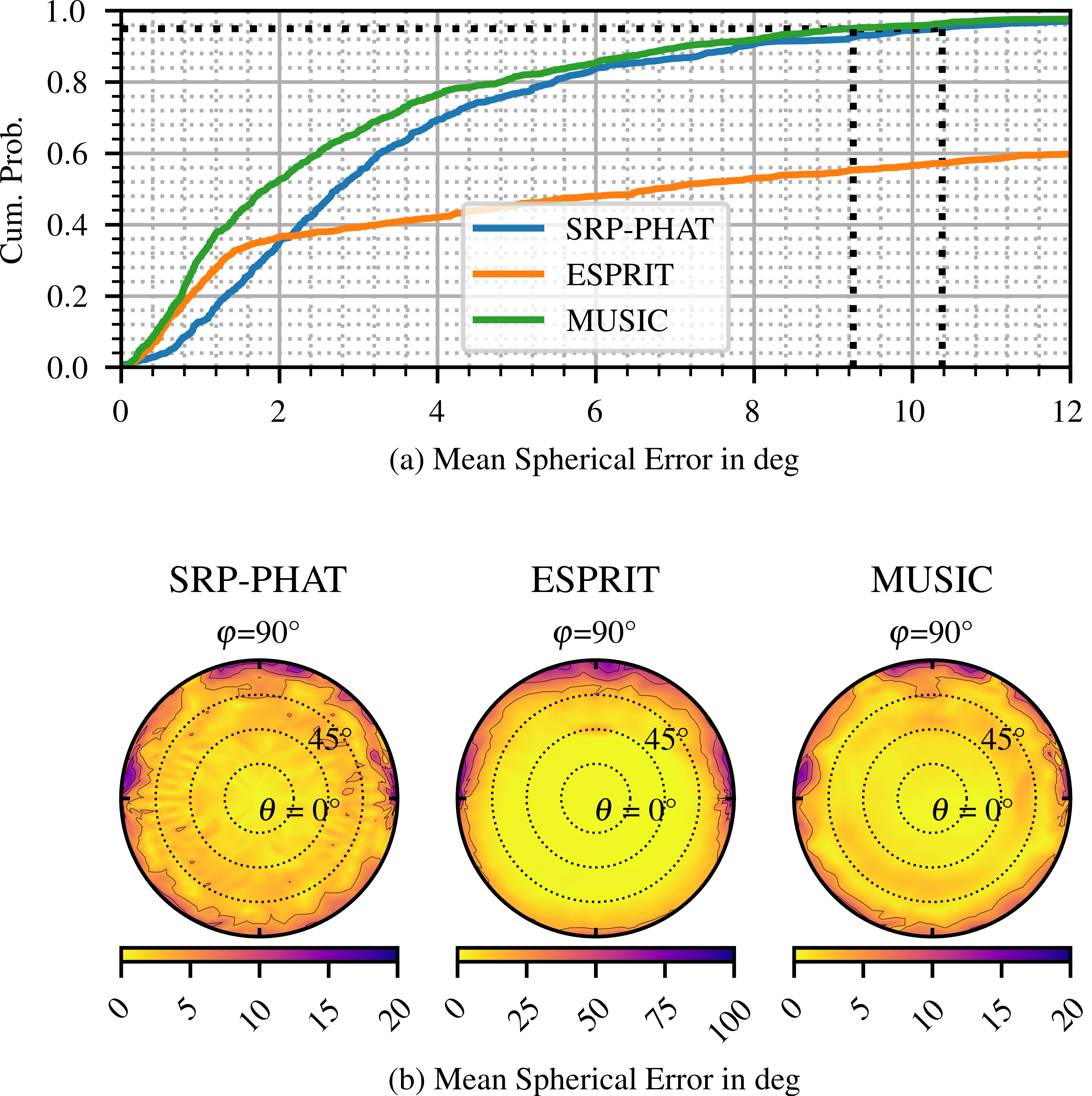}
    \else
        \includesvg[width=\linewidth]{figures/experiments/narrowband-angle-legacy-algorithms-comparison.svg}
    \fi
	\caption{Algorithm evaluation on the URA: (a) presents the cumulative error distribution for the different algorithms. Notably, the MUSIC algorithm achieves the best performance with a 95\% confidence interval error of \SI{9.3}{\degree}. In (b), the error distribution across the measured angle space is depicted. Note the different scaling used for the colors representing the error of the ESPRIT algorithm.}
	\label{fig:result_exp_angle_over_algos}
\end{figure}

Three algorithms are compared using measured data: SRP-PHAT \cite{DiBiase.2000}, Unitary ESPRIT \cite{Zoltowski.1996}, and MUSIC \cite{Schmidt.1986}. A grid comprising 1036 different azimuth and elevation pairs is sampled with the array. The collected samples are preprocessed with a chunk size of $N=4096$, after which the different estimators are applied to the processed data.
In \autoref{fig:result_exp_angle_over_algos} (a), the overall resulting cumulative distribution of the measured spherical error, calculated using \autoref{eq:spherical-error}, is depicted. MUSIC performs the best with \SI{9.3}{\degree} in the 95th percentile, followed by SRP-PHAT with \SI{10.4}{\degree}.
\autoref{fig:result_exp_angle_over_algos} (b) illustrates the distribution of errors, with the elevation angle on the radial axis. The largest spherical error occurs in the high elevation angle regions, primarily due to inaccuracies in the elevation angle estimation. Notably, towards the broadside of the array, the error becomes minimal for all investigated algorithms.
Note the different color scale used for the ESPRIT algorithm.

\subsubsection{Error Distribution}
\begin{figure}
	\centering
		\resizebox{\linewidth}{!}{\input{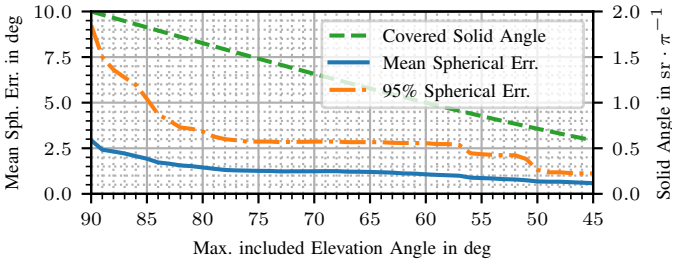}}
	\caption{Error distribution evaluation: To assess error distribution, certain points above a threshold are excluded from the evaluation. On the secondary axis, the available solid angle in normalized steradians $\mathrm{sr} / \pi$ is depicted, serving as a reference for understanding the extent of the \ac{FoV} lost by excluding specific angle sets.}
	\label{fig:angle-exclusion}
\end{figure}

As depicted in \autoref{fig:result_exp_angle_over_algos}, it is evident that the largest errors occur in the lower elevation regions. When excluding these points, thereby narrowing the field of view, the attainable error gets minimized. \autoref{fig:angle-exclusion}  displays the mean and 95th percentile error alongside the left field of view in steradians when excluding data above a certain elevation angle. The steep negative slope of the 95th percentile curve illustrates this phenomenon. For instance, by excluding all angles above 75 degrees, the attainable 95th percentile error reduces to \SI{2.8}{\degree} (or \SI{1.26}{\degree} mean spherical error) while still retaining around \SI{74}{\percent} of the \ac{FoV}. In a real application, this can be implemented by ensuring overlapping regions of multiple receiver arrays.

\subsubsection{Comparison of sparse array geometries}
In order to compare sparse array geometries to each other, we used the same data and parameters as in the previous section. Each sparse array's data is then constructed by masking only the microphones which are present in the current geometry under evaluation and thus, only the data of those microphones is included. Spatial smoothing is performed, using the apertures in x- and y-direction of the difference co-array of each respective geometry as window sizes. Finally, the 2-D Unitary \ac{ESPRIT} algorithm is applied on the spatially smoothed sample covariance matrix to estimate the \ac{DoA} of the sound source. In addition to this, the \ac{URA} and Random arrays are included as a reference. For the URA, once with spatial smoothing as discussed, and once without spatial smoothing with the estimation algorithm directly working with the standard sample covariance matrix.

To evaluate geometries capable of estimating more sources than there are sensors, the unitary ESPRIT estimator with spatial smoothing is essential. This methodology is specifically applicable to geometries that feature a coherent \ac{URA} centerpiece in their difference co-array. This criterion is met by the geometries discussed herein, with the exception of the random array. Given the prior introduction of the random array, it is included in the analysis using the standard MUSIC algorithm, which does not have the capability to estimate more sources than sensors. Interestingly, a comparison of the 95th percentile error, as illustrated in \autoref{fig:angle-exclusion}, reveals a mere increase of approximately \SI{1}{\degree} in error compared to the fully populated \ac{URA}.

\begin{figure*}
	\centering
        
    \if\isPrint1
        \includegraphics[width=\linewidth]{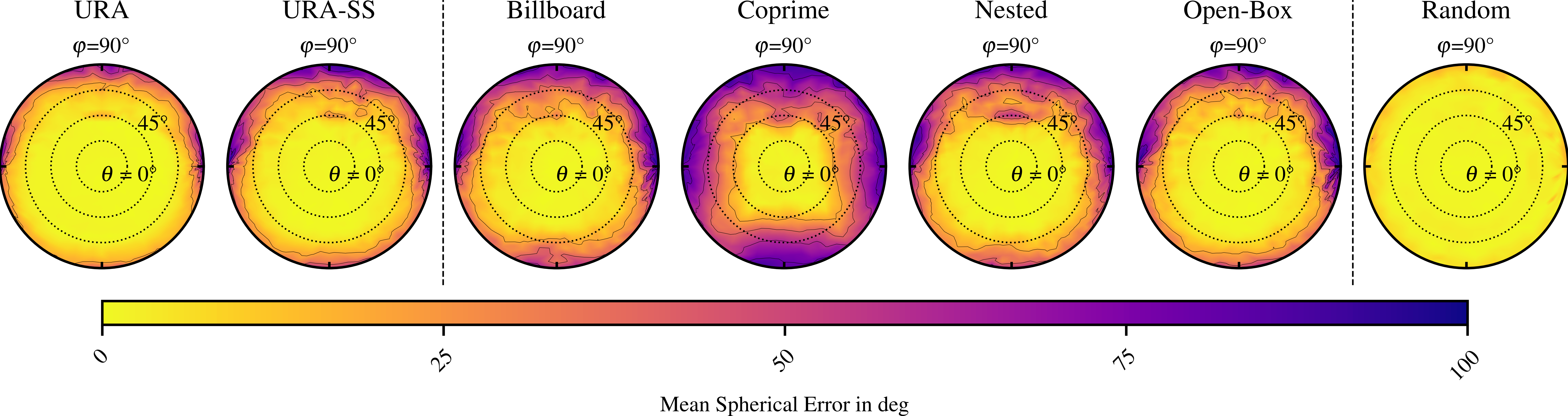}
    \else
        \includesvg[width=\linewidth]{figures/experiments/narrowband-angle-legacy-geometries-spherical.svg}
    \fi
    \caption{Error distribution analysis of different array geometries reveals a similar trend as in the algorithms evaluation, with errors accumulating in the high elevation angle regions. It's noteworthy that the MUSIC estimator is utilized for the Random array. Conversely, the ESPRIT algorithm with Spatial Smoothing is employed for the rest of the geometries. Additionally, spatial smoothing is applied for URA-SS, a necessity for the other sparse geometries to achieve data augmentation.}
	\label{fig:result_exp_angle_over_sparse_geometries_esprit}
\end{figure*}

\begin{figure}[!htbp]
	\begin{centering}
		\input{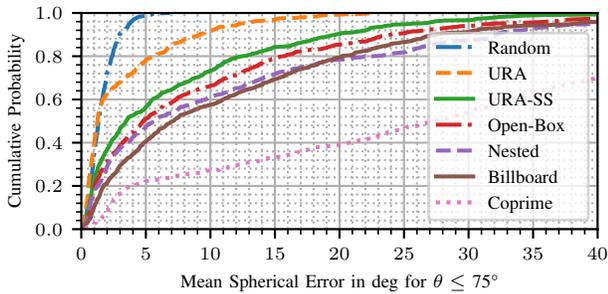}
		\caption{The cumulative distribution function of the mean spherical error is computed with higher elevations $\theta\leq\SI{75}{\degree}$ excluded.}
		\label{fig:sparse-geometries-cdf}
	\end{centering}
\end{figure}

\autoref{fig:result_exp_angle_over_sparse_geometries_esprit} illustrates the error distribution for the different array geometries. Consistent with previous observations, errors tend to accumulate in the high elevation angle regions across all geometries. Notably, the Coprime array exhibits a strong azimuth angle dependency in the error distribution, corroborating findings from \autoref{fig:array-geometries-beampattern}. Conversely, the other sparse geometries demonstrate comparable performance, as indicated by \autoref{fig:sparse-geometries-cdf}, which displays the \ac{CDF}. In this evaluation, the Open-Box array performs best within the category of sparse geometries, followed by the Nested and Billboard arrays (cf. \autoref{tab:array-geometries-eva-results}). As expected, the Coprime array, characterized by the smallest aperture size and fewest number of sensors, demonstrates the poorest performance.

\begin{table}
    \centering
    \renewcommand{\arraystretch}{1.2}%
    \begin{tabular}{@{}lccc@{}}\toprule
         Array & \thead{50\% CI in deg} & \thead{95\% CI in deg} & \thead{Estimator}  \\
         \midrule
        URA & 1.26 & 12.22 & U. ESPRIT \\
        URA-SS & 3.49 & 25.57 & U. ESPRIT + SS  \\
        \midrule
        Billboard & 7.17 & 38.24 & U. ESPRIT + SS  \\
        Coprime & 27.45 &  62.96 & U. ESPRIT + SS  \\
        Nested & 5.74 & 39.41 & U. ESPRIT + SS  \\
        Open-Box & \textbf{4.85} &  \textbf{31.85} & U. ESPRIT + SS  \\
        \midrule
        Random &  1.34 & 3.76 & MUSIC  \\
        \bottomrule
    \end{tabular}
    \caption{Distribution of the spherical error across different array geometries with 50- and 95-percentile \acp{CI}. This data only takes elevation angles with $\theta\le\SI{75}{\degree}$ into account.}
    \label{tab:array-geometries-eva-results}
\end{table}

\subsection{DoA-Estimation (Multiple Sources)}
\begin{figure}
	\begin{centering}
		\resizebox{\linewidth}{!}{\input{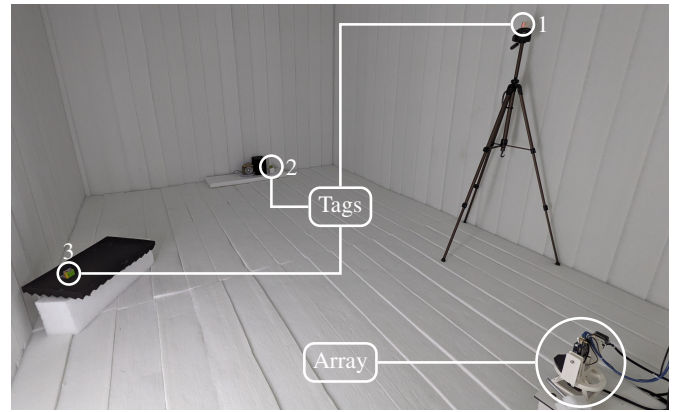}}
		\caption{The multisource experiment setup involves positioning three tags within an anechoic chamber at varying angles. Each tag reproduces a distinct Walsh-Hadamard sequence. The signals emitted by the tags are sampled by the array located in the lower right-hand corner.}
		\label{fig:anechoic_chamber_two_sources}
	\end{centering}
\end{figure}

\begin{figure}
	\centering
 
    \if\isPrint1
        \includegraphics[width=\linewidth]{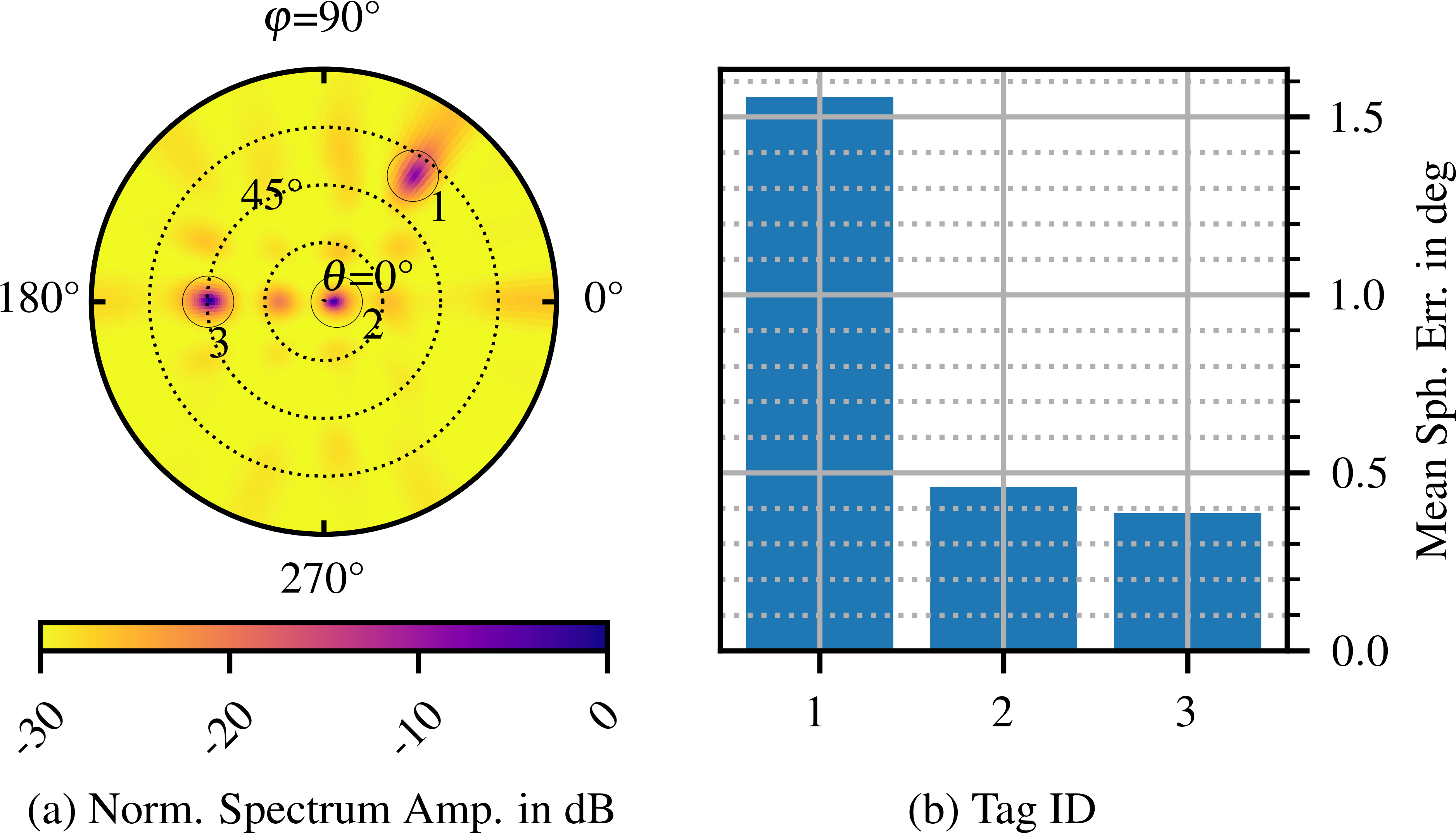}
    \else
        \includesvg[width=\linewidth]{figures/experiments/multiple-sources-calc-composite.svg}
    \fi
    \caption{Multiple source estimation using the Nested array and the standard MUSIC estimator instead of a data-augmented covariance matrix estimator. In (a), the MUSIC pseudospectrum of the scene is depicted, with different tag IDs highlighted. (b) illustrates the mean spherical error for the various tag positions.}
	\label{fig:multisource_results}
\end{figure}

In this experiment, three cooperative sources (Tags) are positioned within the anechoic chamber using a laser which is mounted on a two-axis rotation table, as depicted in \autoref{fig:anechoic_chamber_two_sources}. The tags are equipped with piezoelectric speakers. The source signals, which are bandlimited ($B=\SI{200}{\hertz}$) Walsh-Hadamard sequences mixed to \SI{20}{\kilo\hertz}, ensure good cross-correlation properties. The tags emit signals simultaneously at an incident sound pressure ranging from \SI{50}{\decibel~SPL} (Tag 2) to \SI{55}{\decibel~SPL} (Tag 1). Instead of employing the data-augmented ESPRIT estimator, the MUSIC estimator is applied to the nested array selection of the captured data, as depicted in Figure \ref{fig:hw_concept_array}. \autoref{fig:multisource_results} presents the MUSIC pseudospectrum of the captured scene along with the mean spherical error for all three sources. The error ranges from \SI{0.4}{\degree} up to \SI{1.6}{\degree}, primarily depending on the elevation angle.


\subsection{Discussion}
In the initial experiments, we observed that calibration provided only marginal benefits, particularly in lower elevation angle regions, highlighting the robustness of the hardware concept and the quality of the used microphones. Subsequent analysis of single-source experiments revealed that errors tend to accumulate in higher elevation angle regions. When designing localization systems, it may be advisable to ensure overlapping regions where angle measurements in the higher elevation angle range can be discarded or de-weighted.
The \ac{URA}, while not isotropic, does not exhibit strong, noticeable azimuth angle-dependent behavior. Furthermore, we observed that the general-purpose SRP-PHAT algorithm, widely applied in various applications, performed worse than the specialized MUSIC algorithm, especially for narrowband signals. Comparison of multiple array geometries unveiled their performance on real-world data, with some geometries struggling with azimuth angle dependency, resulting in inferior estimation performance in general.
Notably, increasing sensor count and aperture size typically leads to better performance. However, distributing sensors too sparsely in various directions to optimize aperture size can yield azimuthal dependencies. It's also worth mentioning that estimating multiple concurrent sources is possible by employing orthogonal codes, which may also enable source identification.

\section{Conclusions and Outlook}
In this contribution, a measurement platform has been proposed for the evaluation of several sparse array geometries. The platform consists of 64 microphones in URA configuration, tuned to a maximum frequency of around \SI{20.8}{\kilo\hertz}.  Through a series of simulations and experiments, we have demonstrated the feasibility and effectiveness of sparse array geometries in accurately estimating the direction of arrival of sound sources in indoor environments.
The developed array demonstrates promising performance, achieving a 95-percentile spherical error of \SI{2.8}{\degree} (or a mean error of \SI{1.26}{\degree}) when high-error producing regions are excluded using the MUSIC estimator.
Our investigation encompassed various sparse array geometries, revealing notable angle dependencies in some configurations. Among these, geometries like the Open-Box or Nested array exhibited superior performance on applied metrics.
Furthermore, an experiment involving three concurrently emitting sources demonstrated the feasibility of resolving multiple sources simultaneously.

In future research, we aim to explore the capability of estimating more sources than sensors. Key steps towards this objective will involve finding optimized orthogonal codes to ensure optimal cross-correlation properties.

\bibliographystyle{IEEEtran}
\bibliography{literature,literature-georg}

\begin{IEEEbiography}[{\includegraphics[width=1in,height=1.25in,clip,keepaspectratio]{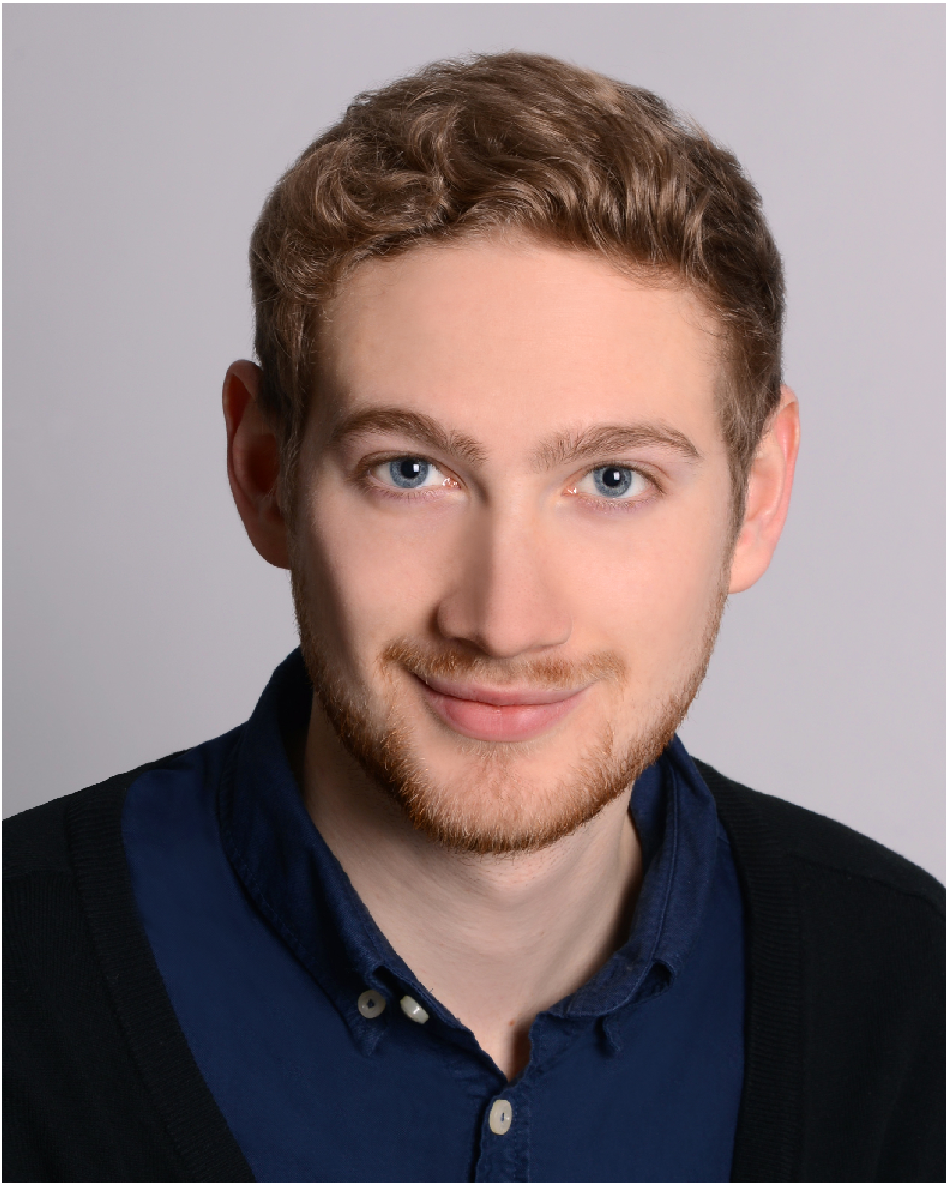}}]{Georg K.J. Fischer}
    received the B.Sc.{} degree in electrical engineering from the TU Braunschweig, Germany in 2017 and the M.Sc.{} degree in electrical engineering and information technology from the Karlsruhe Institute of Technology (KIT), Germany in 2019.
    He joined the Fraunhofer Ernst-Mach-Institute (EMI) in 2020, as a research assistant in the field of signal processing.
\end{IEEEbiography}
\vskip -2\baselineskip plus -1fil

\begin{IEEEbiography}[{\includegraphics[width=1in,height=1.25in,clip,keepaspectratio]{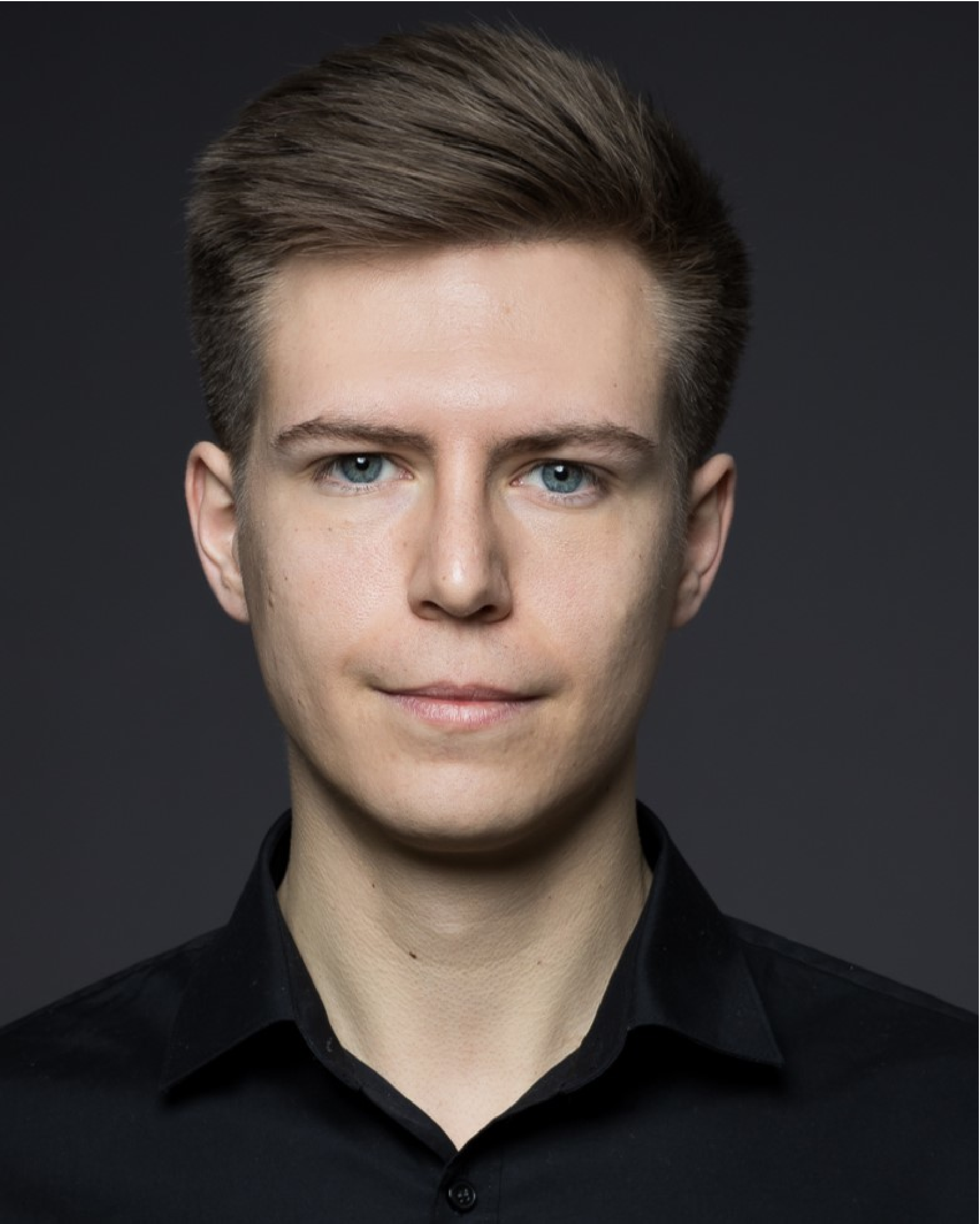}}]{Niklas Thiedecke}
    received the B.Sc. degree in electrical engineering and information technology from the TU Kaiserslautern, Germany in 2018 and the M.Sc. degree in embedded systems engineering in 2022 from the University of Freiburg, Germany.
    He was employed from 2018 to 2019 at Bosch Sensortec GmbH, Germany, where he participated in Bosch’s PreMaster-program, working in the field of smart sensor development. Since August 2022, he’s employed at Bosch Sensortec GmbH, Germany as a system architect for pre-development and development of smart sensors and smart connected sensors.
\end{IEEEbiography}
\vskip -2\baselineskip plus -1fil

\begin{IEEEbiography}[{\includegraphics[width=1in,height=1.25in,clip,keepaspectratio]{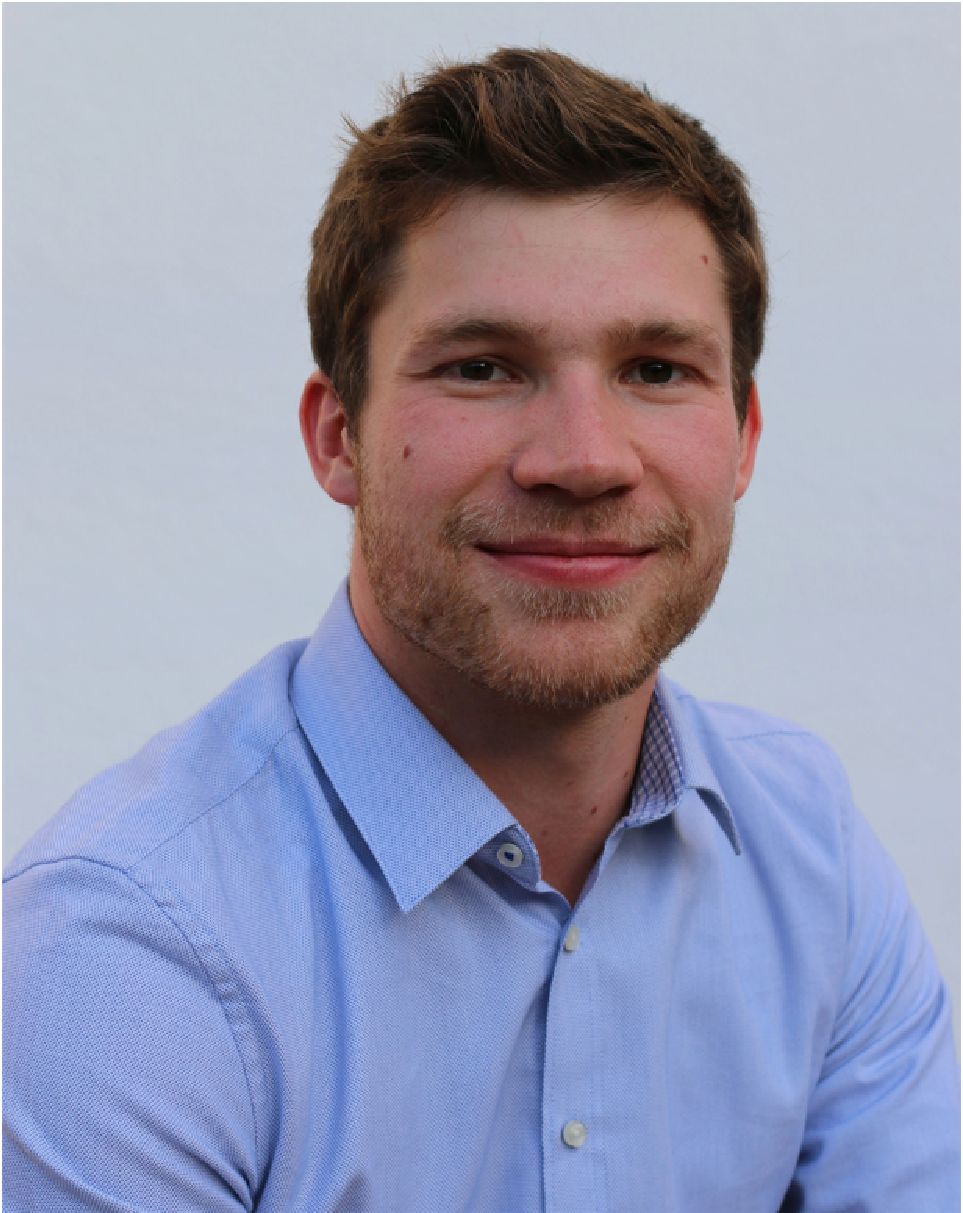}}]{Thomas Schaechtle}
 received his B.Eng. degree in electrical engineering from the Cooperative State University DHBW Lörrach in 2013 with TDK-Micronas, Freiburg, as education partner and the M.Sc. degree in embedded systems engineering in 2018 from the University of Freiburg. In 2018, he started as a Ph.D candidate at the Laboratory of Electrical Instrumentation and Embedded Systems, University of Freiburg. He was employed from 2010 to 2014 at TDK-Micronas, where he worked one year as an application engineer in a gas sensor research group. Since April 2022, he has been also working as research assistant for the Fraunhofer Ernst-Mach-Institute (EMI) in the field of ultrasonic communication.
\end{IEEEbiography}
\vskip -2\baselineskip plus -1fil

\begin{IEEEbiography}[{\includegraphics[width=1in,height=1.25in,clip,keepaspectratio]{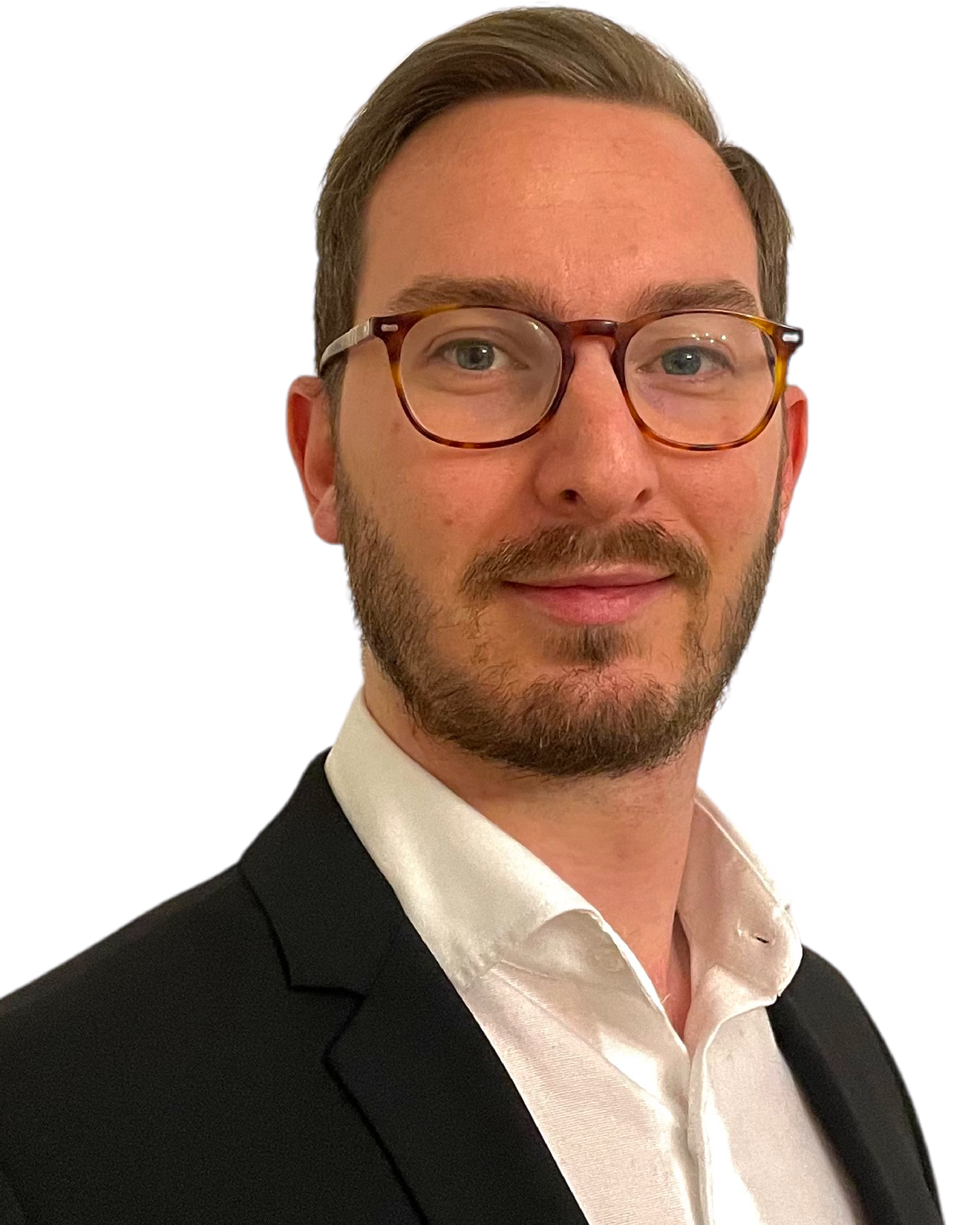}}]{Andrea~Gabbrielli}
    received the B.Sc. degree in telecommunications engineering from the Faculty of Engineering of the University of Pisa, Pisa, Italy, in 2014, the M.Sc. degree in computer science and the Ph.D. degree in microsystems engineering from the Faculty of Engineering of the University of Freiburg, Freiburg im Breisgau, Germany, in 2019 and 2024, respectively. He joined the Laboratory for Electrical Instrumentation and Embedded Systems at the Department of Microsystems Engineering (IMTEK), University of Freiburg, Germany, in 2019, as a scientific researcher in the field of signal processing for indoor localization.
\end{IEEEbiography}
\vskip -2\baselineskip plus -1fil

\begin{IEEEbiography}[{\includegraphics[width=1in,height=1.25in,clip,keepaspectratio]{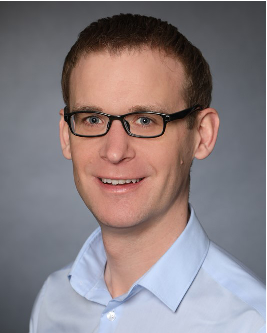}}]{Fabian H{\"o}flinger}
    received the B.Sc.{} degree in automation engineering from the University of Applied Sciences Ravensburg-Weingarten, Weingarten, Germany, in 2007, the master's degree in automation and energy systems from the Mannheim University of Applied Sciences, Mannheim, Germany, in 2007, and the Ph.D. degree in microsystems engineering, University of Freiburg, Freiburg im Breisgau, Germany, in 2014, with a focus on localization systems.
    		
    He was with Junghans Feinwerktechnik, Dunningen/Seedorf, Germany, where he developed components for telemetric systems. From 2007 to 2010, he was a Development Engineer. Since 2010, he has been with the Laboratory for Electrical Instrumentation and Embedded Systems, Department of Microsystems Engineering, University of Freiburg, where he has also been a Group Leader since 2014 and has a research group in the field of indoor localization. Since 2019, he has also been working for the Fraunhofer Ernst-Mach-Institute (EMI).
\end{IEEEbiography}
\vskip -2\baselineskip plus -1fil

\begin{IEEEbiography}[{\includegraphics[width=1in,height=1.25in,clip,keepaspectratio]{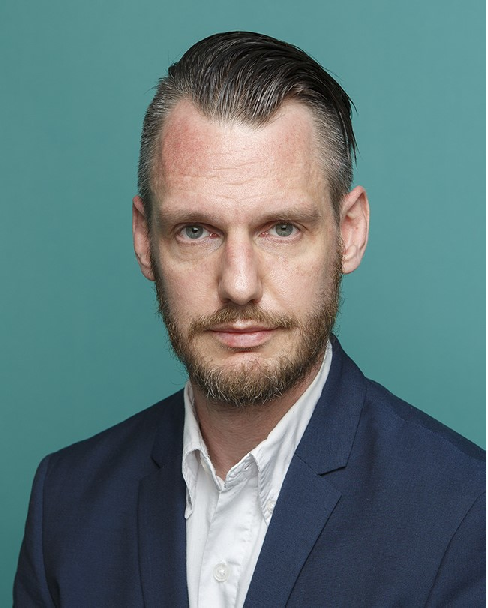}}]{Alexander Stolz}
    holds the Professorship for Resilience Engineering of Technical Systems at the Department of Sustainable Systems Engineering at the Faculty of Engineering of the Albert-Ludwigs-University in Freiburg.
    He studied civil engineering at University of Wuppertal and wrote his doctoral thesis about mobilization of bedding stresses in granular soil at the Professorship for Geotechnique. As head of the department Safety, Security and Resilience of Technical Systems he is specialized in the experimental investigation and numerical modeling of complicated and complex structures and networks under extraordinary disruptive events. 
    Dr. Stolz coordinated the Thematic Group “Resistance of structures to explosion effects” in the ERNCIP (European Reference network for Critical Infrastructure protection) framework from 2012 till 2020 and is since 2017 also appointed international member of the AMR10 Committee on Critical Infrastructure Protection.

\end{IEEEbiography}
\vskip -2\baselineskip plus -1fil

\begin{IEEEbiography}[{\includegraphics[width=1in,height=1.25in,clip,keepaspectratio]{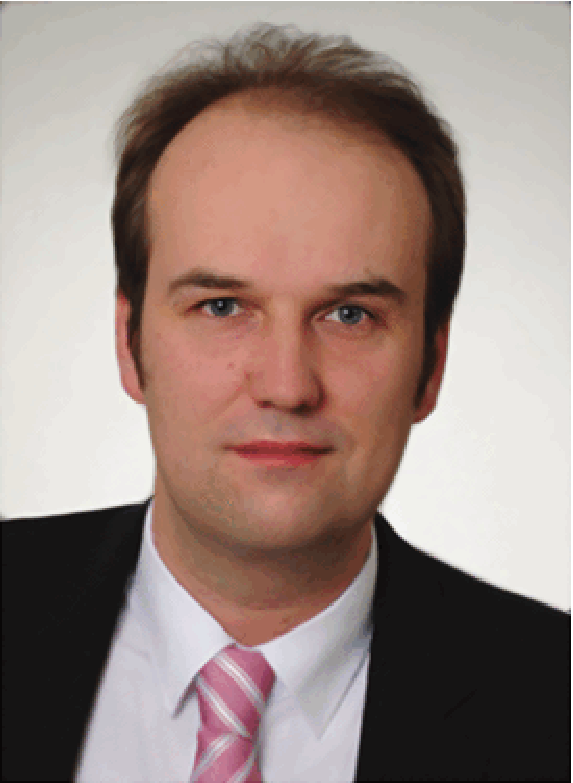}}]{Stefan J. Rupitsch}
   (Member, IEEE) was born in Kitzbuehel, Austria, in 1978. He received the Diploma and Ph.D. degrees in mechatronics from Johannes Kepler University Linz, Austria, in 2004 and 2008, respectively, and the Habilitation degree from the University of Erlangen–Nuernberg, Germany, in 2018.
   
   In 2004, he was a Junior Researcher with the Linz Center of Mechatronics, Linz. From 2005 to 2008, he was with the Institute for Measurement Technology, Johannes Kepler University Linz. From 2008 to 2020, he was with the Chair of Sensor Technology, University of Erlangen–Nuernberg, where he held a Deputy Professorship. Since December 2020, he has been a Full Professor for electrical instrumentation and embedded systems with the University of Freiburg.
   
   His research interests include piezoelectric transducers, energy harvesting, embedded systems, ultrasonic imaging and therapy, simulation-based material characterization, and noncontact measurements. He has authored more than 150 articles in these fields as well as the book Piezoelectric Sensors and Actuators: Fundamentals and Applications.
   
   Stefan J. Rupitsch was a recipient of the Austrian Society of Measurement and Automation Technology Award for his Ph.D. dissertation in 2009 and the Outstanding Paper Award of the Information Technology Society in 2016. He is an Associate Editor of the IEEE SENSORS JOURNAL and the TM-Technisches Messen journal. He serves as a Guest Associate Editor for the Journal of Sensors and Sensor Systems (JSSS).
\end{IEEEbiography}

\vskip -2\baselineskip plus -1fil

\end{document}